\documentclass[11pt,graphicx]{article}
\usepackage[brazil,USenglish]{babel}
\usepackage[utf8]{inputenc}
\usepackage[T1]{fontenc}
\usepackage{indentfirst}
\usepackage{latexsym}
\usepackage{amsmath,amssymb,amsfonts,wasysym}
\usepackage{graphicx}
\usepackage{epstopdf} 
\usepackage[left=3cm,right=2cm,top=3cm,bottom=3cm]{geometry} 
\usepackage[center]{caption} 
\usepackage[pdftex, colorlinks=true, linkcolor=black, citecolor=black, urlcolor=black]{hyperref}
\usepackage[all]{hypcap}
\usepackage{fancyhdr}
\usepackage[Conny]{fncychap} 
\nonstopmode
\usepackage{setspace}
\usepackage{braket}

\begin{document}
\title{Casimir Effect in Horava-Lifshitz-like theories}
\author{I. J. Morales Ulion,  
 E. R. Bezerra de Mello \thanks{E-mail: emello@fisica.ufpb.br} 
and A. Yu. Petrov \thanks{Email: petrov@fisica.ufpb.br}\\
Dept. de Física-CCEN. Universidade Federal da Paraíba \\
58.059-970, J. Pessoa, PB. C. Postal 5.008. Brazil}
\maketitle
\begin{abstract}

In this paper we consider a Lorentz-breaking scalar field theory within the Horava-Lifshtz 
approach. We investigate the changes that a space-time anisotropy produces in the Casimir 
effect. A massless real quantum scalar field is considered in two distinct situations: 
between two parallel plates and inside a rectangular two-dimensional box. In both cases we 
have adopted specific boundary conditions on the field at the boundary. As we shall see, 
the energy and the Casimir force strongly depends on the parameter associated with the 
breaking of Lorentz symmetry and also on the boundary conditions.

\end{abstract}

\bigskip


\bigskip

\bigskip


\bigskip

\newpage
\renewcommand{\thesection}{\arabic{section}.}
\section{Introduction}

A free quantum field theory (QFT) can be treated as an infinite quantum system of simple harmonic oscillators, 
with its fundamental excitations interpreted as the associated particles. Thus, the vacuum in QFT is the state 
in which all quantum oscillators  are in its ground state. But, as we know, the energy of the ground state of a 
quantum harmonic oscillator is not zero. Consequently, the vacuum energy,  being the sum of the energies of the ground 
states of these oscillators, is infinite. The QFT offers several examples  which show that this vacuum plays 
a fundamental role not only in the physics of microscopic phenomena, but also the physics of 
macroscopic phenomena. One of these phenomena the Casimir effect \cite{Casimir}.

The Casimir effect is one of the most notable consequences of vacuum quantum fluctuations. 
In its most general description, the Casimir effect is a consequence of the changes caused 
in the vacuum energy due to the presence of boundary conditions imposed on the fields. 
The effect was first predicted theoretically by H. B Casimir in 1948 \cite{Casimir}, 
and experimentally confirmed ten years later  by M. J. Sparnnaay \cite{Sparnaay}. 
In the 90s, experiments have confirmed the Casimir effect with high degree of accuracy \cite{Lamoureux,Moh}. 
In his original work Casimir predicted that due to quantum fluctuations of the electromagnetic field, 
two parallel flat neutral (grounded) plates attract each other with a force given by:
\begin{equation}
F = - A \frac{\pi^2 \hbar c}{240 a^4}\   ,
\end{equation}
where $A$ is the area of plates and $a$ is the distance between them. The Casimir effect 
is traditionally studied by changing up the idealized effects of borders by boundary conditions.

Once the theory of relativity is the basis for QFT, the Lorentz symmetry is a fully conserved symmetry 
in this theory. However, other theories include models where the Lorentz symmetry is  violated. 
In the quantum gravity, Ho\u rava-Lifshitz (HL) theory is a theory where the Lorentz symmetry is 
broken in a strong manner. The space-time anisotropy caused by the breaking of Lorentz symmetry, 
occurs due to different properties of scales in which coordinates space and time are set, so that 
the theory is invariant under the rescaling $x\to bx$, $t\to b^{\xi}t$, where $\xi$ is a number 
called the critical exponent \cite{Horava}. The space-time anisotropy in a given field 
theory model should certainly modify the spectrum of the Hamiltonian operator given.

The HL approach, or, as is the same, the idea of the space-time anisotropy, clearly can be 
applied not only to gravity, but also to other field theory models, including scalar, spinor 
and gauge theories. One of the main reasons for interest to HL-like generalizations of these 
theories consists in possible improvement of convergence of quantum corrections. Among the most 
important results achieved within their studies, one can emphasize calculation of the one-loop 
effective potential in HL-like QED and HL-like Yukawa model \cite{EP,Far,Lima} and study of different 
issues related to renormalization of these theories \cite{ren,Iengo,Gomes,12}. Therefore, the problem of study 
of the Casimir effect in HL-like generalizations of different field theories seems to be quite 
natural. Some preliminary studies in this direction, for a very particular case, were performed in \cite{Petrov}.

In this paper we intend to generalize the results obtained in \cite{Petrov} 
through the analysis the influence of the combination of the two above 
mentioned effects on the vacuum energy associated with a real quantum scalar field, i.e.,
by the imposition of specific boundary conditions on the fields and the Lorentz-breaking symmetry. 
In fact, we want to study the quantum scalar field in two different configurations:
between two parallel plates and inside a rectangular two-dimensional box. In section \ref{Sect2}
we briefly derive the equation of motion obeyed by the field in a Lorentz symmetry breaking 
context. In section \ref{Sect3} we  explicitly develop the calculations considering the
quantum system confined between two parallel plates and a two-dimensional rectangular box.
In both situations we impose on the field Dirichlet, Neumann and mixed boundary conditions on the
boundaries. Finally we leave for Conclusions \ref{Concl} our most relevant remarks found in this paper.

\section{Klein-Gordon equation with the Lorentz Symmetry Breaking}
\label{Sect2}

The Lorentz invariance, known as a cornerstone of the quantum field theory, began to be intensively questioned
in recent decades, both within theoretical and experimental  contexts. In 1989, V. A. Kostelecky and S. 
Samuel \cite{KSamuel} described a mechanism in string theory which allows violation of Lorentz 
symmetry at the Planck energy scale. This mechanism is based on a spontaneous breaking of the Lorentz 
symmetry implemented through acquiring non-zero vacuum expectation values by some vectors 
or tensors, which implies privileged directions and hence anisotropy in space-time. This effect 
is called {\it condensation} of tensors in vacuum. The breaking happened at the Planck energy 
scale in a more fundamental theory, and the effects of this break manifest themselves for other energy scales in 
different field theory models, for example the Standard Model, were not detected up to now, 
because such effects are suppressed by powers of the Planck mass. Kostelecky and Samuel 
also evaluated that this idea of Lorentz symmetry breaking should be incorporated into 
the Standard Model (SM), thus giving rise to the Standard Model Extended (SME). 
The proposed Lorentz symmetry breaking is intensively tested through observations and experiments. 
So, astronomic observations in the star spectrum, show an evidence that the fine structure constant 
$\left( \alpha = \frac{e^2}{\bar{h} c}\right)$, which is a measure of intensity of the electromagnetic 
interaction between photons and electrons, slowly varies \cite{Songaila,Dav,Songaila2}.
Later studies indicated that other mechanisms for breaking of Lorentz symmetry are also possible, 
such as space-time noncommutativity  \cite{Carroll,Anis,Carl,Hew,Bert}, variation of coupling constants \cite{Kost,Bert2,Bert3} and modifications of quantum gravity  \cite{Alfaro,Alfaro2}.

Further, in the paper \cite{Anselm1,Anselm2}, the concept of large Lorentz symmetry breaking, 
or, as it is mostly called, space-time asymmetry, has been introduced. Following this idea, 
the time and space coordinates, and derivatives with respect to them, enter in the action 
in different degrees, so, the action continues to be quadratic in time derivatives
to avoid arising the ghosts, but involves higher spatial derivatives, 
whose order is $2\xi$. Namely these theories will be the main object of our study here.

\subsection{Modified Klein-Gordon Equation}

In the seminal work \cite{Horava} Ho\u rava called a great attention to theories with space-time 
asymmetry, since it revitalized a hope to solve the key problem of quantum gravity, that is, 
to find a renormalizable and ghost-free gravity theory. Indeed,
the four-dimensional HL gravity is power-counting renormalizable for $\xi=3$.

Using the HL approach, in the next section we will discuss how the Lorentz symmetry  
violation interferes in the vacuum structure of a quantum scalar field theory. We 
will study the Casimir effect which has been well studied in the usual field theories. 
Before that, we first need to see how the HL theory modifies the Klein-Gordon equation.

We work with a theory which space-time coordinates no longer have the same {\it weight}, 
as in the case where Lorentz symmetry is preserved. In this scenario, we consider 
the theory of a massless real scalar field, which is the simplest case. The action
associated with this system is given by \cite{Anselm1,Anselm2}:
\begin{equation}
\label{S}
S = \frac{1}{2} \int \mathrm{d}t \mathrm{d}^d x \left( \partial_{0}\phi\partial_{0}\phi
 - l^{2(\xi-1)}\partial_{i_{1}}\partial_{i_{2}}...\partial_{i_{\xi}}
\phi\partial_{i_{1}}\partial_{i_{2}}...\partial_{i_{\xi}}\phi \right) \  .
\end{equation}

In the case (3 + 1)-dimensions the equation of motion  is:
\begin{equation}
\label{kgmodific}
[\partial_{0}^2 +  l^{2(\xi -1)}(-1)^{\xi} 
\partial_{i_{1}}...\partial_{i_{\xi}}\partial_{i_{1}}...\partial_{i_{\xi}}]\phi = 0  \   .
\end{equation}

The equation \eqref{kgmodific}  is the modified Klein-Gordon equation within the HL-like theory. 
It will be used in next section where we will study the changes that the breaking of Lorentz 
symmetry implies on the Casimir effect.

\section{Violation of Lorentz Symmetry within the Casimir Effect}
\label{Sect3}

In this section we will study how the space-time anisotropy generated by 
the HL theory modifies the results of the Casimir effect associated with a
massless scalar quantum field. In what follows, we study two distinct cases. 
In the first one we consider this effect between two parallel plates and the 
second inside a two-dimensional rectangular box. In both cases we deal with 
three kinds of boundary conditions.

\subsection{Parallel Plates}

We consider a massless scalar field inside two parallel plates, as it is shown in figure \ref{fig1}. 
As we saw in the previous section, the equation that a massless real scalar field  must satisfy 
in the HL theory is given by:
\begin{figure}[h]
\centering
\includegraphics[height=5cm, width=8cm]{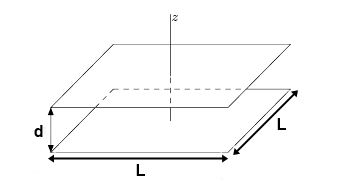}
\caption{Two parallel plates with area $L^2$ separated by a distance $d<<L$.}
\label{fig1}
\end{figure}
\begin{equation}
\label{kgmodific1}
[\partial_{0}^2 +  l^{2(\xi -1)}(-1)^{\xi} \partial_{i_{1}}...
\partial_{i_{\xi}}\partial_{i_{1}}...\partial_{i_{\xi}}]\phi(x) = 0 \   . 
\end{equation}

First we must obtain the solution for \eqref{kgmodific1} by imposing on the fields 
specific boundary conditions and thus obtain the Hamiltonian $\hat{H}$ operator.
Afterwords, we can calculate the total vacuum energy of the system and then 
determine the Casimir energy. Subsequently we will obtain Casimir force. In the 
$(3 + 1)$-dimensional case, the term  $\partial_{i_{1}}...
\partial_{i_{\xi}}\partial_{i_{1}}...\partial_{i_{\xi}}$ takes the following form:
\begin{equation}
\partial_{i_{1}}...\partial_{i_{\xi}}\partial_{i_{1}}...\partial_{i_{\xi}} 
= (\partial_{x}^2+ \partial_{y}^2+ \partial_{z}^2)^{\xi} \  ,
\end{equation}
so the equation for $\phi(x)$ reads,
\begin{equation}
\label{eqmovql}
\left[\partial_{0}^2 +  l^{2(\xi -1)}(-1)^{\xi} (\partial_{x}^2+ \partial_{y}^2+ \partial_{z}^2)^{\xi}\right]\phi(x) = 0.
\end{equation}

\subsubsection*{Dirichlet Condition}

Now we must solve \eqref{eqmovql} requiring that the solution must satisfies 
the Dirichlet boundary conditions given below,
\begin{equation}
\phi(x)_{z=0} =\phi(x)_{z=d} = 0 \  . 
\end{equation}

Adopting the standard procedure \cite{Mandl}, we write the field operator as
\begin{equation}
\label{ocdh}
\hat{\phi}(x) = \int \mathrm{d}^2\textbf{k} \sum_{n=1}^{\infty} 
\frac{1}{[(2\pi)^2 d\, k_{0}]^{1/2}} \sin\left(\dfrac{n\pi}
{d}z\right)[a_{\textbf{k},n}e^{-i k x} + a^{\dagger}_{\textbf{k},n}e^{i k x}] \  ,
\end{equation}
where $a_{\textbf{k},n}$ and $a^{\dagger}_{\textbf{k},n}$ correspond to the 
annihilation and creation operators, respectively, characterized by the set of quantum numbers
$\sigma=\{k_x, k_y, n\}$. These operators satisfy the algebra 
\begin{equation}
\label{algebra}
\begin{split}
[a_{\textbf{k},n}, a^{\dagger}_{\textbf{k}',n'}] & = \delta_{n,n'}\delta^2(\textbf{k}-\textbf{k}') , \\
[a_{\textbf{k},n}, a_{\textbf{k}',n'}] = & [a^{\dagger}_{\textbf{k},n}, a^{\dagger}_{\textbf{k}',n'}] = 0 \  .
\end{split}
\end{equation}
In \eqref{ocdh} we have defined $kx \equiv k_{0}x_{0} - k_{x}x - k_{y}y$, being
\begin{equation}
k_{0} = l^{\xi-1}\omega_{\textbf{k},n}^{\xi} \ ,
\end{equation}
with $\omega_{\textbf{k},n}$ obeying the dispersion relation,
\begin{equation}
\omega_{\textbf{k},n}^2 = k_{x}^2 + k_{y}^2 + \left(\frac{n\pi}{d}\right)^2 \ .
\end{equation}

The Hamiltonian operator, $\hat{H}$, for this case is given by:
\begin{equation}
\hat{H} = \frac{l^{\xi-1}}{2}\int \mathrm{d}^2\textbf{k} \sum_{n=1}^{\infty} 
\omega_{\textbf{k},n}^{\xi}\left[2 a^{\dagger}_{\textbf{k,n}} a_{\textbf{k,n}} + 
\frac{L^2}{(2\pi)^2}\right] \   .
\end{equation}
Consequently the vacuum energy is obtained by taking the vacuum expectation value of $\hat{H}$:
\begin{equation}
\label{vacuum}
E_{0} = \bra{0}\hat{H}\ket{0} = \frac{l^{\xi-1} L^2}{8 \pi^2}\int \mathrm{d}^2\textbf{k} 
\sum_{n=1}^{\infty} \omega_{\textbf{k},n}^{\xi} \  .
\end{equation}

In order to develop the summation on the quantum number $n$, we shall use the Abel-Plana formula \cite{Adv.Cas,Ford},
\begin{equation}
\label{Abel}
\sum_{n=0}^{\infty}F(n) = \frac{1}{2}F(0) + \int_{0}^{\infty} F(t)\mathrm{d}t + 
i \int_{0}^{\infty}\frac{\mathrm{d}t}{e^{2\pi t}-1}[ F(it) - F(-i t)] \  . 
\end{equation}

Performing in \eqref{vacuum} a change of coordinates in the plane $(k_{x},k_{y})$ 
to polar coordinate, we get
\begin{equation}
\label{vacuum1}
E_{0} = \frac{l^{\xi-1} L^2}{4\pi}\int_{0}^{\infty} 
\mathrm{d}k k \left[- \frac{1}{2}F(0) + \int_{0}^{\infty} 
F(t)\mathrm{d}t + i\int_{0}^{\infty} \frac{F(it) - F(-it)}{e^{2 \pi t} - 1}\mathrm{d}t\right] \   ,
\end{equation}
where 
\begin{equation}
F(n) = \left[k^2 + \left(\frac{n\pi}{d}\right)^2\right]^{\xi/2}.
\end{equation}
Note that the first term on the right-hand side of \eqref{vacuum1} 
refers to vacuum energy in the presence of  only one plate, and the
second one is connected with vacuum energy without boundary. Both
terms do not contribute to the Casimir energy. As a
result, the Casimir energy per unit area of the planes is given by
\begin{equation}
\label{E-C0}
\frac{E_{C}}{L^2} = i \frac{l^{\xi-1}}{4\pi}\int_{0}^{\infty} 
\mathrm{d}k	k \int_{0}^{\infty}\mathrm{d}t \frac{[k^2 + 
(\frac{i t \pi}{d})^2]^{\xi/2} - [k^2 + (-\frac{i t \pi}{d})^2]^{\xi/2}}{e^{2 \pi t} - 1} \  .
\end{equation}
Performing a change of variable, where $\frac{t\pi}{d} = u$, we get
\begin{equation}
\frac{E_{C}}{L^2} = i \frac{l^{\xi-1}  d}{4\pi^2}\int_{0}^{\infty} 
\mathrm{d}k	k \int_{0}^{\infty}\mathrm{d}u \frac{[k^2 + (i u)^2]^{\xi/2} - 
[k^2 + (-i u)^2]^{\xi/2}}{e^{2 d u} - 1}.
\end{equation}

The integral over the {$u$} variable  must be considered in two cases, 
for \underline{$k>u$} and \underline{$k<u$}. Thus integrating can provide two different values.
\begin{itemize}
\item \underline{For $k>u$}:
\end{itemize}
In this range we have,
\begin{equation}
\label{k>u}
[k^2 + (\pm i u)^2]^{\xi/2} = [k^2 - u^2]^{\xi/2}.
\end{equation}
\begin{itemize}
\item \underline{For $k<u$}:
\end{itemize}
In this range we have, 
\begin{equation}
\label{k<u}
\begin{split}
[k^2 + (\pm i u)^2]^{\xi/2} = e^{\pm i\xi \frac{\pi}{2}} [u^2 - k^2]^{\xi/2}.
\end{split}
\end{equation}

Consequently the integral $u$ in the interval $[0,k]$ vanishes. So, we get:
\begin{equation}
\begin{split}
\frac{E_{C}}{L^2}  &= i \frac{l^{\xi-1} d}{4\pi^2}\int_{0}^{\infty} 
\mathrm{d}k	k \int_{k}^{\infty}\mathrm{d}u \frac{(u^2-k^2)^{\xi/2}}{e^{2 d u} - 1} 
(e^{i \xi \frac{\pi}{2}} - e^{ -i \xi \frac{\pi}{2}}), \\ \\ &=  - 
\sin\left(\frac{\pi \xi}{2}\right) \frac{l^{\xi-1} L^2 d }
{2 \pi^2}\int_{0}^{\infty} \mathrm{d}k	k \int_{k}^{\infty}\mathrm{d}u 
\frac{(u^2-k^2)^{\xi/2}}{e^{2 d u} - 1}.
\end{split}
\end{equation}

To solve the integral in $u$ we introduced a new variable  $t=u/k$, so
\begin{equation}
\frac{E_{C}}{L^2} = - \sin\left(\frac{\pi \xi}{2}\right) 
\frac{l^{\xi-1} d }{2 \pi^2}\int_{1}^{\infty}(t^2 - 1)^{\xi/2}\mathrm{d}t 
\int_{0}^{\infty}\mathrm{d}k \frac{k^{\xi + 2}}{e^{2 d k t} - 1}\   .
\end{equation}

Using the expression below \cite{Grad},
\begin{equation}
\int_{0}^{\infty}\mathrm{d}x \frac{x^{\nu - 1}}{e^{\mu x} -1} = \frac{1}{\mu^{\nu}}\Gamma(\mu)\zeta(\mu),
\end{equation} 
where $\Gamma(\mu)$ and $\zeta(\mu)$ are the gamma and the Riemann zeta functions, respectively, we get,
\begin{equation}
\frac{E_{C}}{L^2} = - \sin\left(\frac{\pi \xi}{2}\right) \frac{l^{\xi-1} d }
{2 \pi^2} \frac{\Gamma(\xi+3)\zeta(\xi+3)}{(2 d)^{\xi+3}}\int_{1}^{\infty}\mathrm{d}t\frac{(t^2 - 1)^{\xi/2}}{t^{\xi+3}}.
\end{equation}

Therefore Casimir energy per unit area is expressed by:
\begin{equation}
\label{E-C}
\frac{E_{C}}{L^2} = - \sin\left(\frac{\pi \xi}{2}\right) \frac{ l^{\xi-1}  
 \Gamma(\xi+2)\zeta(\xi+3)}{2^{\xi+4} \hspace{0.2cm} \pi^2 d^{(\xi+2)}} \   .
\end{equation}
From \eqref{E-C} the Casimir pressure between two 
parallel planes due to scalar field oscillations takes the
form,
\begin{equation}
P(d) = - \sin\left(\frac{\pi \xi}{2}\right) \frac{ l^{\xi-1}   
\Gamma(\xi+3)\zeta(\xi+3)}{2^{\xi+4} \hspace{0.2cm} \pi^2 d^{(\xi+3)}} \   .
\end{equation}

An important remarks about the above result is that the Casimir pressure 
depends on the critical exponent through the term $\sin(\frac{\pi \xi}{2})$. 
For even value of this exponent the pressure vanishes; however, for odd values
it changes its sign. It can be positive, which corresponds to a repulsive force, 
or negative, which corresponds to an attractive force. 

For $\xi=1$, the Casimir pressure reproduces the usual one
given in \cite{Petrov}:
\begin{equation}
\label{C-Press}
P(d) = - \frac{\pi^2}{480 d^4}.
\end{equation}

\subsubsection*{Neumann Condition}

Now we want to obtain solutions of Eq. \eqref{eqmovql} which obey the
boundary condition below,
\begin{equation}
\frac{\partial \phi(\textbf{x})}{\partial z}\Big |_{z=0} = 
\frac{\partial \phi(\textbf{x})}{\partial z}\Big |_{z=d} = 0.
\end{equation}

After some intermediate steps, we can say that for this case the field operator reads, 
\begin{equation}
\label{ocnh}
\hat{\phi}(x) = \int \mathrm{d}^2\textbf{k} \sum_{n=0}^{\infty} c_{n} 
\cos\left(\dfrac{n\pi}{d}z\right)[a_{\textbf{k},n}e^{-i k x} + a^{\dagger}_{\textbf{k},n}e^{i k x}] \  ,
\end{equation}
where
\begin{equation}
c_{n} = \begin{cases}
\dfrac{1}{\sqrt{2(2\pi)^2 d k_{0}}} & \text{for $n = 0$}, \\
\dfrac{1}{\sqrt{(2\pi)^2 d k_{0}}} & \text{for $n \ge 0$},
\end{cases}
\end{equation}
and $k_{0} = l^{\xi-1}\omega^{\xi}_{\textbf{k},n}$. Here $\omega_{\textbf{k},n}$ 
obeys the same dispersion relation as in the previous case:

\begin{equation}
\omega_{\textbf{k},n}^2 = k_{x}^2 + k_{y}^2 + \left(\frac{n \pi}{d}\right)^2.
\end{equation}

Using field operator and the commutation relation \eqref{algebra}, the operator 
$\hat{H}$ takes the form
\begin{equation}
\hat{H} = \frac{l^{\xi-1}}{2}\int\mathrm{d}^2\textbf{k}
\sum_{n=0}^{\infty}\text{'} \omega_{\textbf{k},n}^{\xi} 
\left[2 a^{\dagger}_{\textbf{k},n} a_{\textbf{k},n} + \frac{L^2}{(2\pi)^2}\right] \  ,
\end{equation}
where the prime in the summation symbol means that the term with $n=0$ should be
divided by two. 

The vacuum energy is given by
\begin{equation}
E_{0} = \bra{0}\hat{H}\ket{0} = \frac{l^{\xi-1} L^2}{8 \pi^2} 
\int\mathrm{d}^2\textbf{k}\sum_{n=0}^{\infty}\text{'} \omega_{\textbf{k},n}^{\xi} \,\ .
\end{equation}

Using again the Abel-Plana summation formula and rewriting the integral  
on the plane $(k_{x},k_{y})$ in polar coordinates, we obtain
\begin{equation}
\label{E-C1}
E_{0} = \frac{l^{\xi-1} L^2}{4 \pi} \int_{0}^{\infty}\mathrm{d}k k 
\left[\int_{0}^{\infty}F(t)\mathrm{d}t + i \int_{0}^{\infty} \mathrm{d}t \frac{F(it) - F(-it)}{e^{2 \pi t} - 1}\right],
\end{equation}
where here
\begin{eqnarray}
F(n) = \omega_{\textbf{k},n}^{\xi} = \left[k^2 + \big(\frac{n\pi}{a}\big)^2\right]^{\xi/2} \  . 
\end{eqnarray}

As in previous cases, the Casimir energy per unit
area is given by the second term on the right side of \eqref{E-C1}. So, we get
\begin{equation}
\frac{E_{C}}{L^2} = i \frac{l^{\xi-1}}{4 \pi} \int_{0}^{\infty}
\mathrm{d}k k\int_{0}^{\infty}\mathrm{d}t 
\frac{[k^2 + (\frac{i t \pi}{d})^2]^{\xi/2} - [k^2 + (-\frac{i t \pi}{d})^2]^{\xi/2}}{e^{2 \pi t} - 1}.
\end{equation}

This is the same expression obtained for the Dirichlet case, Eq. \eqref{E-C0}.
Consequently the Casimir energy per unit area is given by:
\begin{equation}
\frac{E_{C}}{L^2} = - \sin\left(\frac{\pi \xi}{2}\right) \frac{ l^{\xi-1}  
 \Gamma(\xi+2)\zeta(\xi+3)}{2^{\xi+4} \hspace{0.2cm} \pi^2 d^{(\xi+2)}} \ .
\end{equation}
The Casimir pressure reads,
\begin{equation}
P(d)= - \sin\left(\frac{\pi \xi}{2}\right) \frac{ l^{\xi-1}   
\Gamma(\xi+3)\zeta(\xi+3)}{2^{\xi+4} \hspace{0.2cm} \pi^2 d^{(\xi+3)}}.
\end{equation}
As in the previous case, the Casimir pressure is zero for a even $\xi$ and
change its signal for odd values of this parameter.

\subsubsection*{Mixed Condition}

Now let us consider a scalar field which obeys a Dirichlet boundary condition on one plane,
and a Neumann boundary condition on the other. Two different configurations take place:
\begin{itemize}
\item First configuration,
\begin{equation}
\phi(z=0) = \frac{\partial\phi(\textbf{x})}{\partial z} |_{z=d}=0 \ .
\end{equation}
\item Second configuration,
\begin{equation}
\frac{\partial\phi(\textbf{x})}{\partial z} |_{z=0} = \phi(z=d)  =0 \ .
\end{equation}
\end{itemize}
After solving Eq. \eqref{eqmovql} with these conditions, the obtained fields operators 
can be shown to look like:
\begin{equation}
\hat{\phi}_{(a)}(x) = \int \mathrm{d}^2 \textbf{k} \sum_{n = 0}^{\infty} \frac{1}
{\sqrt{(2 \pi)^2 d k_{0}}} \sin\left((n+1/2)\frac{ \pi}{d}z\right)
[ a_{\textbf{k},n}e^{-ikx} + a_{\textbf{k},n}^{\dagger} e^{ikx}]
\end{equation}
for the first configuration and
\begin{equation}
\hat{\phi}_{(b)}(x) = \int \mathrm{d}^2 \textbf{k} \sum_{n = 0}^{\infty} 
\frac{1}{\sqrt{(2 \pi)^2 d k_{0}}} \cos\left((n+1/2)\frac{ \pi}
{d}z\right)[ a_{\textbf{k},n}e^{-ikx} + a_{\textbf{k},n}^{\dagger} e^{ikx}] \  , 
\end{equation}
for the second configuration. In both cases $k_{0}= l^{\xi-1}\omega_{\textbf{k},n}^{\xi}$ 
and $\omega_{\textbf{k},n}$ satisfies the dispersion relation,
\[\omega_{\textbf{k},n}^{2} = k_{x}^2 + k_{y}^2 + \big[(n+1/2)\frac{\pi}{d}\big]^2.\]

Both field operators, $\hat{\phi}{^a}(x)$ and $\hat{\phi}^{b}(x)$, provide the same 
Hamiltonian operator,
\begin{equation}
\hat{H} = \frac{l^{\xi-1}}{2}\int\mathrm{d}^2\textbf{k}\sum_{n=0}^{\infty} 
\omega_{\textbf{k},n}^{\xi} \left[2 a^{\dagger}_{\textbf{k},n} a_{\textbf{k},n} + 
\frac{L^2}{(2\pi)^2}\right] \   .
\end{equation}

The vacuum energy of the scalar field is expressed as
\begin{equation}
E_{0} = \bra{0}\hat{H}\ket{0} = \frac{l^{\xi-1} L^2}{8 \pi^2} 
\int\mathrm{d}^2\textbf{k}\sum_{n=0}^{\infty} \omega_{\textbf{k},n}^{\xi} .
\end{equation}

Changing the coordinates of the plane $(k_ {x},k_ {y})$ to polar ones, 
and using the Abel-Plana  summation formula for half-integer numbers \cite{Adv.Cas}, we get:
\begin{equation}
E_{0} = \frac{l^{\xi-1} L^2}{4 \pi} \int_{0}^{\infty}\mathrm{d}k k \left\lbrace 
\int_{0}^{\infty}F(t)\mathrm{d}t - i \int_{0}^{\infty}\mathrm{d}t\frac{F(it) - 
F(-it)}{e^{2 \pi t} + 1}\right\rbrace  \  .
\end{equation}
Again, the Casimir energy is given by the second term of the above expression. 
The first one refers to the free energy vacuum. Then the Casimir energy is given by
\begin{equation}
E_{C} = - i \frac{l^{\xi-1} L^2}{4 \pi} \int_{0}^{\infty}\mathrm{d}k k 
\int_{0}^{\infty}\mathrm{d}t\frac{[k^2 + \left(\frac{i t \pi}{d}\right)^2]^{\xi/2} 
+ [k^2 + \left(-\frac{i t \pi}{d}\right)^2]^{\xi/2}}{e^{2 \pi t} + 1} \   .
\end{equation}
After performing a change of variable, $\frac{t\pi}{d}=u$, the Casimir energy by unit area reads,
\begin{equation}
\label{E-C-M}
\frac{E_{C}}{L^2} = - i \frac{l^{\xi-1} d}{4 \pi^2} 
\int_{0}^{\infty}\mathrm{d}k k \int_{0}^{\infty}\mathrm{d}u
\frac{[k^2 + (iu)^2]^{\xi/2} + [k^2 +(-iu)^2]^{\xi/2}}{e^{2 a u} + 1}.
\end{equation}

Again, we must consider the integral over the variable $u$ in two sub-intervals:  
The first one is $[0, k]$ and the second is $[k, \infty)$.  
From  \eqref{k>u} we have to the integral in the interval $[0,k]$ vanishes, 
so it remains only the integral in the second
interval. By using \eqref{k<u}, we get: 
\begin{eqnarray}
\frac{E_{C}}{L^2} &=& - i \frac{l^{\xi-1}  d}{4\pi^2}\int_{0}^{\infty} 
\mathrm{d}k	k \int_{k}^{\infty}\mathrm{d}u \frac{(u^2-k^2)^{\xi/2}}{e^{2 d u} + 1}
 (e^{i \xi \frac{\pi}{2}} - e^{ -i \xi \frac{\pi}{2}})\  , \nonumber\\
 &=&{\mathrm{sin}}\left(\frac{\pi \xi}{2}\right) \frac{l^{\xi-1} d }{2 \pi^2}\int_{0}^{\infty} 
\mathrm{d}k	k \int_{k}^{\infty}\mathrm{d}u \frac{(u^2-k^2)^{\xi/2}}{e^{2 d u} + 1}\  .
\end{eqnarray}

Introducing a new variable $t$, where $u=kt$, we get
\begin{equation}
E_{C} = {\mathrm{sin}}\left(\frac{\pi \xi}{2}\right) 
\frac{l^{\xi-1} L^2 d }{2 \pi^2} \int_{1}^{\infty}\mathrm{d}t (t^2 - 1)^{\xi/2} 
\int_{0}^{\infty} \mathrm{d}k	\frac{k^{\xi+2}}{e^{2 d k t} + 1} \   .
\end{equation}

Using \cite{Grad} one finds that the Casimir energy per unit area is given by
\begin{equation}
\frac{E_{C}}{L^2} = {\mathrm{sin}}\left(\frac{\pi \xi}{2}\right) 
(1-2^{-(\xi+2)})\frac{l^{\xi-1} \Gamma(\xi+2)\zeta(\xi+3)}{2^{\xi+4}\hspace{0.2cm}
 \pi^2 \hspace{0.2cm} d^{\xi+2} } \  .
\end{equation}
For this case, the Casimir pressure reads,
\begin{equation}
\frac{F_{C}}{L^2} = (1-2^{-(\xi+2)}) {\mathrm{sin}}\left(\frac{\pi \xi}{2}\right) 
\frac{l^{\xi-1} \Gamma(\xi+3)\zeta(\xi+3)}{2^{\xi+4}\hspace{0.2cm} \pi^2 \hspace{0.2cm} d^{\xi+3} }.
\end{equation}

We again have here the same general behavior: for $\xi$ even the Casimir force vanishes, 
while for $\xi$ odd the Casimir force switches between an attractive and repulsive one. 
We can also see that for $\xi=$1 this force in fact recovers the usual Casimir pressure,
\begin{equation}
P(d) = \frac{7}{3840}\frac{\pi^2}{d^4} \ .
\end{equation}
This result differs from \eqref{C-Press} by the factor a numerical factor and 
the sign. 

\subsection{The Casimir Effect In Rectangular Boxes}

\begin{figure}[h]
\centering
\includegraphics[height=6cm, width=10cm]{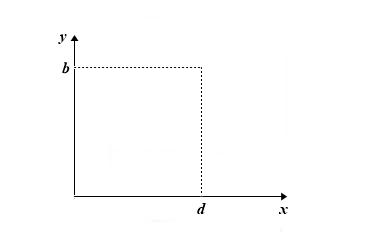}
\caption{Rectangular boxe, with edges $b$ and $d$.}
\label{fig2}
\end{figure}

Now we will consider a massless scalar quantum field, $\phi(x)$, within a two-dimensional 
rectangular boxes defined by $0 \le x \le d$ and $0\le y \le b$, as shown in figure \ref{fig2}.
We will disregard the $z$ coordinate, since it does not cause any 
influence on the Casimir energy. Thus, the field equation $\phi(x)$ must satisfy is
\begin{equation}
\label{emret}
[\partial_{0}^2 + l^{2(\xi-1)}(-1)^\xi (\partial_{x}^2 + \partial_{y}^2)^\xi]\phi(x) = 0 \   .
\end{equation}

Again, in this section we are interested to calculate the Casimir
energy and Casimir force by imposing three different boundary 
conditions on the field as shown below.

\subsubsection*{Dirichlet}

First let us obtain the solution of Eq. \eqref{emret} satisfying the condition below:
\begin{equation}
\begin{cases} 
\phi(t,0,y) = \phi(t,d,y) = 0\  ,\\
\phi(t,x,0) = \phi(t,x,b) = 0\   .
\end{cases}
\end{equation}

The resulting field operator, $\hat{\phi}(x)$ compatible with these condition is:

\begin{equation}
\label{phi2}
\hat{\phi}(x) = \sum_{n,m=1}^{\infty} \sqrt{\frac{2}{d b k_{0}}} 
\sin\left(\frac{n \pi}{d} x\right) \sin\left(\frac{n \pi}{b} y\right)
[a_{n,m}e^{-i k_{0}t} + a^{\dagger}_{n,m}e^{i k_{0}t}] \  ,
\end{equation}
where $k_{0} = l^{\xi-1} \omega^{\xi}_{n,m}$ and $\omega_{n,m}$ obey the dispersion relation,
\begin{equation}
\omega_{n,m} =\sqrt{ \left(\frac{n\pi}{d}\right)^2 + \left(\frac{m\pi}{b}\right)^2} \ .
\end{equation}
In \eqref{phi2} $a_{n,m}$ and $a^{\dagger}_{n,m}$ are the annihilation and
creation operators respectively satisfying the following commutation relations:
\begin{equation}
\label{rcr}
\begin{split}
&[a_{n,m},a^{\dagger}_{n',m'}] = \delta_{n,n'}\delta_{m,m'} \ , \\
&[a_{n,m},a_{n',m'}] = [a^{\dagger}_{n,m},a^{\dagger}_{n',m'}] = 0 \ .
\end{split}
\end{equation}

Thus the Hamiltonian operator $\hat{H}$ for this case is given by
\begin{equation}
\hat{H} = l^{\xi-1} \sum_{n,m=1}^{\infty} \omega^{\xi}_{n,m}
\left[a^{\dagger}_{n,m}a_{n,m} +\frac{1}{2}\right] \  ,
\end{equation}

So the vacuum energy is:
\begin{equation}
\label{ECDiric}
E_{0} = \bra{0}\hat{H}\ket{0} = \frac{l^{\xi-1}}{2}\sum_{n,m=1}^{\infty} 
\omega^{\xi}_{n,m}\  .
\end{equation}

To develop the sums over the quantum numbers $n$ and $m$ we will again use the Abel-Plana 
summation formula separately. First we will develop the sum over $m$ considering $F(m) = [\left(\frac{n\pi}{d}\right)^2 + 
\left(\frac{m\pi}{b}\right)^2]^{\xi/2}$. Doing that we get:
\begin{equation}
\begin{split}
E_{0} = \frac{l^{\xi-1}}{2}\sum_{n=1}^{\infty} \Big[ &	
-\frac{1}{2}\left(\frac{n\pi}{d}\right)^{\xi} + \int_{0}^{\infty}
\mathrm{d}t\left[\left(\frac{n\pi}{d}\right)^2 + \left(\frac{t\pi}{b}\right)^2\right]^{\xi/2} + \\ 
& i \int_{0}^{\infty}\mathrm{d}t\frac{[\left(\frac{n\pi}{d}\right)^2 + 
\left(\frac{it\pi}{b}\right)^2]^{\xi/2} - \left(\frac{n\pi}{d}\right)^2 +
 \left(\frac{-it\pi}{b}\right)^2]^{\xi/2}}{e^{2 \pi t}-1}\Big]\  .
\end{split}
\end{equation}

The last integral should be divided in two parts, as shown below:
\begin{itemize}
\item For $\frac{t}{b}<\frac{n}{d}$ we have:
\end{itemize}
\begin{equation}
\label{sum1}
\left[\left(\frac{n\pi}{d}\right)^2 + \left(\pm i\frac{t\pi}{b}\right)^2\right]^{\xi/2} 
= \left[\left(\frac{n\pi}{d}\right)^2 - \left(\frac{t\pi}{b}\right)^2\right]^{\xi/2}\  .
\end{equation}
\begin{itemize}
\item For $\frac{t}{b}>\frac{n}{d}$ we have:
\end{itemize}
\begin{equation}
\label{sum2}
\left[\left(\frac{n\pi}{d}\right)^2 + \left(\pm i\frac{t\pi}{b}\right)^2\right]^{\xi/2}
 = e^{\pm i \xi \frac{\pi}{2}}\left[\left(\frac{t\pi}{b}\right)^2 - 
\left(\frac{n\pi}{d}\right)^2\right]^{\xi/2} \ .
\end{equation}

Then, from \eqref{sum1} and \eqref{sum2} we have
\begin{equation}
\begin{split}
\label{E-C-box}
E_{0} = \frac{l^{\xi-1}}{2}\sum_{n=1}^{\infty} \Big[ & 
\underbrace{-\frac{1}{2}\left(\frac{n\pi}{d}\right)^{\xi}}_{I} + 
\underbrace{\int_{0}^{\infty}\mathrm{d}t\left[\left(\frac{n\pi}{d}\right)^2 + 
\left(\frac{t\pi}{b}\right)^2\right]^{\xi/2}}_{II} -\\ &  \underbrace{2  
\sin(\xi\frac{\pi}{2})\int_{nb/d}^{\infty}\mathrm{d}t\frac{[\left(\frac{t\pi}{b}\right)^2 - 
\left(\frac{n\pi}{d}\right)^2]^{\xi/2} }{e^{2 \pi t}-1}}_{III} \Big].
\end{split}
\end{equation}

Let us perform the sum in $n$ proceeding with each term separately. Applying the Abel-
Plana formula in the term $I$ of \eqref{E-C-box}, we find	
\begin{eqnarray}
\label{I}
I&=& -\frac{1}{2}\left\lbrace \int_{0}^{\infty}\mathrm{d}v 
\left(\frac{v \pi}{d}\right)^\xi + i\int_{0}^{\infty}\mathrm{d}v 
\frac{(\frac{i v \pi}{d})^\xi - (-\frac{i v \pi}{d})^\xi}{e^{2 \pi v}-1} \right\rbrace \nonumber\\
&=& -\frac{1}{2}\int_{0}^{\infty}\mathrm{d}v \left(\frac{v \pi}{d}\right)^\xi
 + \sin\left(\xi\frac{\pi}{2}\right) \left(\frac{\pi}{d}\right)^\xi \int_{0}^{\infty}\mathrm{d}v \frac{v^\xi}{e^{2 \pi v}-1}.
\end{eqnarray}

The integral of the second term of \eqref{I} is obtained from \cite{Grad}. 
Then, the term $I$ from \eqref{E-C-box} is given by:
\begin{equation}
-\frac{1}{2}\sum_{n=1}^{\infty} \left(\frac{n\pi}{d}\right)^{\xi}  = 
-\frac{1}{2}\int_{0}^{\infty}\mathrm{d}v \left(\frac{v \pi}{d}\right)^\xi + 
\sin\left(\xi\frac{\pi}{2}\right) \left(\frac{\pi}{d}\right)^\xi \frac{\Gamma(\xi+1)\zeta(\xi+1)}{(2 \pi)^{\xi+1}}.
\end{equation}

Performing the sum for the term $II$ of  \eqref{E-C-box} we have:
\begin{equation}
\begin{split}
II =  \int_{0}^{\infty}\mathrm{d}t \Bigg\{&  -\frac{1}{2}\left(\frac{t\pi}{b}\right)^\xi 
+  \int_{0}^{\infty}\mathrm{d}v \left[\left(\frac{v\pi}{d}\right)^2 + 
\left(\frac{t\pi}{b}\right)^2\right]^{\xi/2}  \\ &
- i \int_{0}^{\infty}\mathrm{d}v \frac{\left[\left(\frac{i v\pi}{d}\right)^2 + 
\left(\frac{t\pi}{b}\right)^2\right]^{\xi/2} - \left[\left(\frac{- i v\pi}{d}\right)^2 + 
\left(\frac{t\pi}{b}\right)^2\right]^{\xi/2} }{e^{2 \pi v}-1}\Bigg\}.
\end{split}
\end{equation}

The last term of this equation must also be split into two parts, 
for $v<\frac{ta}{b}$ and $v>\frac{ta}{b}$, thus we have:
\begin{eqnarray}
II&= & -\frac{1}{2}\left(\frac{\pi}{b}\right)^\xi\int_{0}^{\infty}t^\xi\mathrm{d}t + 
\int_{0}^{\infty}\mathrm{d}t\int_{0}^{\infty}\mathrm{d}v \left[\left(\frac{v\pi}{d}\right)^2 + 
\left(\frac{t\pi}{b}\right)^2\right]^{\xi/2} \nonumber\\ 
&-&2 \sin\left(\xi\frac{\pi}{2}\right)\int_{0}^{\infty} dt\int_{td/b}^{\infty}dv
\frac{\left[\left(\frac{v \pi}{d}\right)^2 - \left(\frac{t \pi}{b}\right)^2\right]^{\xi/2}}{e^{2 \pi x}-1} \  . 
\end{eqnarray}

The integral of the last term of $II$ can be obtained by using \cite{Grad}, so that $II$ reads,
\begin{equation}
\begin{split}
II = & -\frac{1}{2}\left(\frac{\pi}{b}\right)^\xi\int_{0}^{\infty}t^\xi\mathrm{d}t 
+ \int_{0}^{\infty}\mathrm{d}t\int_{0}^{\infty}\mathrm{d}v \left[\left(\frac{v\pi}{d}\right)^2 
+ \left(\frac{t\pi}{b}\right)^2\right]^{\xi/2} \\ &- 
 \sin\left(\xi\frac{\pi}{2}\right)\frac{d}{b}\left(\frac{\pi}{b}\right)^\xi 
\frac{\Gamma(\xi+2)\zeta(\xi+2)\mathrm{B}(\frac{1}{2},\frac{\xi}{2}+1)}{(2 \pi d/b)^{\xi+2}},
\end{split}
\end{equation}
where $B(x,y)$ is the beta function \cite{Grad}.

Developing the term $III$ of \eqref{E-C-box}, we have:
\begin{equation}
III = - 2 \sin\left(\xi\frac{\pi}{2}\right)\sum_{n=1}^{\infty} 
\int_{nb/d}^{\infty}\mathrm{d}t \frac{[(\frac{t \pi}{b})^2 - 
(\frac{n\pi}{d})^2]^{\xi/2}}{e^{2 \pi t}-1} \ .
\end{equation}
By using the identity $\sum\limits_{m=1}^{\infty}e^{- n m} = \frac{1}{e^n -1}$, we can rewrite $III$ as:
\begin{equation}
III = - 2 \sin\left(\xi\frac{\pi}{2}\right)\sum_{n,m=1}^{\infty} 
\int_{nb/d}^{\infty}\mathrm{d}t \left[\left(\frac{t \pi}{b}\right)^2 - 
\left(\frac{n\pi}{d}\right)^2\right]^{\xi/2}e^{- 2 \pi t m}.
\end{equation}

Defining a new variable $u=\frac{td}{n b}$, the above expressions is rewritten as:
\begin{equation}
III = - 2 \sin\left(\xi\frac{\pi}{2}\right)\left(\frac{\pi}{d}\right)^\xi 
\left(\frac{b}{d} \sum_{n=1}^{\infty} n^{\xi+1}\sum_{m=1}^{\infty} 
\int_{1}^{\infty}\mathrm{d}u[u^2-1]^{\xi/2} e^{- 2 \pi u n m b/d}\right) \ .
\end{equation}
Finally using the integral representation of the modified Bessel function \cite{Grad}, 
\begin{equation}
K_\nu(z)=\frac{(z/2)^\nu\Gamma(1/2)}{\Gamma(\nu+1/2)}\int_1^\infty dt (t^2-1)^{\nu-1/2}
e^{-zt} 
\end{equation}
the term $III$ is given by
\begin{equation}
III = - 2 \sin\left(\xi\frac{\pi}{2}\right)\left(\frac{\pi}{d}\right)^\xi 
\frac{\Gamma(\frac{\xi}{2}+1)}{\pi^{\frac{\xi+2}{2}}} 
\left(\frac{b}{d}\right)^{\frac{1-\xi}{2}}\sum_{n,m=1}^{\infty} 
\left(\frac{n}{m}\right)^{\frac{\xi+1}{2}}K_{\frac{\xi+1}{2}}(2 \pi n m b/d).
\end{equation}

Therefore, substituting the terms $I$, $II$ and $III$ into \eqref{E-C-box}, we have:
\begin{equation}
\label{E0RD}
\begin{split}
E_{0} = \frac{l^{\xi-1}}{2} & \Bigg\{ -\frac{\pi^\xi}{2}\left(\frac{1}{d^\xi}+
\frac{1}{b^\xi}\right)\int_{0}^{\infty}\mathrm{d}t t^\xi +\sin\left(\xi\frac{\pi}{2}\right) 
\left(\frac{\pi}{d}\right)^\xi \frac{\Gamma(\xi+1)\zeta(\xi+1)}{(2 \pi)^{\xi+1}} + \\
\\
\int_{0}^{\infty}\mathrm{d}t\int_{0}^{\infty}&\mathrm{d}v \left[\left(\frac{v\pi}{d}\right)^2 
+ \left(\frac{t\pi}{b}\right)^2\right]^{\xi/2} - \sin\left(\xi\frac{\pi}{2}\right)
\frac{d}{b}\left(\frac{\pi}{b}\right)^\xi \frac{\Gamma(\xi+2)\zeta(\xi+2)\mathrm{B}
(\frac{1}{2},\frac{\xi}{2}+1)}{(2 \pi d/b)^{\xi+2}} \\ 
 \\
 - 2  \sin\left(\xi\frac{\pi}{2}\right)&\left(\frac{\pi}{d}\right)^\xi 
\frac{\Gamma(\frac{\xi}{2}+1)}{\pi^{\frac{\xi+2}{2}}} \left(\frac{b}{d}\right)^{\frac{1-\xi}{2}}
\sum_{n,m=1}^{\infty} \left(\frac{n}{m}\right)^{\frac{\xi+1}{2}}K_{\frac{\xi+1}{2}}(2 \pi n m b/d)\Bigg\} \  .
\end{split}
\end{equation}

The first term of \eqref{E0RD} is proportional to the perimeter of rectangle with sides $b^\xi$ and 
$d^\xi$.\footnote{In the case $\xi=1$ it is easier to visualize that this energy 
refers to the energy in case where we have a rectangle with sides $b$ and $d$ \cite{Adv.Cas}.}
The third term of \eqref{E0RD} refers to the free vacuum energy of the area bounded 
by rectangle $bd$. Because we are interested to obtain vacuum energy arising due to the 
imposition of boundary conditions on all sides of the rectangle $bd$, we can omit these
terms. This process is equivalent to a renormalization of the above result.
So Casimir energy for this case is given by:
\begin{equation}
\begin{split}
E_{C} = \frac{l^{\xi-1} \pi^\xi }{2} \sin\left(\xi\frac{\pi}{2}\right)\Big
\{& \frac{\Gamma(\xi+1)\zeta(\xi+1)}{d^\xi (2 \pi)^{\xi+1}} - 
\frac{b}{d^{\xi+1} (2\pi)^{\xi+2}}\Gamma(\xi+2)\zeta(\xi+2)\mathrm{B}\Big(1/2,\frac{\xi}{2} +1\Big) \\ 
&- 2 \frac{\Gamma(\frac{\xi}{2}+1)}{d^{\frac{\xi+1}{2}} b^{\frac{\xi-1}{2}} 
\pi^{\frac{\xi}{2}+1}}\sum_{n,m=1}^{\infty} \left(\frac{n}{m}\right)^{\frac{\xi+1}{2}}
K_{\frac{\xi+1}{2}}(2 \pi n m b/d)\Big\} \  .
\end{split}
\end{equation}

We can see that for $\xi$ even the Casimir energy vanishes. However, for odd values of this
parameter, the Casimir energy change the sign. In special cases $\xi=1$ and $\xi=3$ we have:
\begin{itemize}
\item For $\xi=1$
\end{itemize}
\begin{equation}
\label{ECBox1}
E^{\xi=1}_{C} = \frac{\pi}{2}\left\lbrace\frac{1}{24 d} - \frac{b}{8 \pi^2 d^2}\zeta(3) - 
\frac{1}{d \pi}\sum_{n,m=1}^{\infty}\frac{n}{m}K_{1}(2\pi n m b/d)\right\rbrace.
\end{equation}
\begin{itemize}
\item For $\xi=3$
\end{itemize}
\begin{equation}
\label{ECbox2}
E^{\xi=3}_{C} = - \frac{l^{2} \pi^{3}}{2}\left\lbrace\frac{1}{240 d^3} - 
\frac{9 b}{32 \pi^4 d^4}\zeta(5) - \frac{3}{2 b d^2 \pi^2}
\sum_{n,m=1}^{\infty}\frac{n^2}{m^2}K_{2}(2\pi n m b/d)\right\rbrace.
\end{equation}
So, we see that \eqref{ECBox1} reproduces the usual result given in \cite{Adv.Cas}.

From the above results we can obtain the Casimir force acting on the edges of the boxes: 
\begin{itemize}
\item For \underline{$\xi=1$}:
\end{itemize}
\begin{equation}
\begin{split}
F^{1}_{d} = - \frac{\partial E^{\xi=1}_{C}}{\partial d} = -\frac{\pi}{2}
\Bigg\{ -\frac{1}{24 d^2} + \frac{b}{4 \pi^2 d^3}\zeta(3) + 
\frac{1}{\pi d^2}\sum_{n,m=1}^{\infty}\frac{n}{m}K_{1}(2 \pi n m b/d)
 \\ - \frac{1}{\pi d}\sum_{n,m=1}^{\infty}\frac{n}{m}\frac{\partial K_{1}(2 \pi n m b/d)}{\partial d}\Bigg\} 
\end{split}
\end{equation}
is the force acting on the edges located in $x=0$ and $x=d$. 
\begin{equation}
F^{1}_{b} = - \frac{\partial E^{\xi=1}_{C}}{\partial b} = 
-\frac{\pi}{2}\left\lbrace - \frac{\zeta(3)}{8 \pi^2 d^2} - 
\frac{1}{\pi d}\sum_{n,m=1}^{\infty}\frac{n}{m}\frac{\partial 
K_{1}(2 \pi n m b/d)}{\partial b}\right\rbrace  
\end{equation}
is the force acting on the edges in $y=0$ and $y=b$. \\
For both results we have,
\begin{equation}
\frac{\partial K_{1}(F(x))}{\partial x} = -
\frac{1}{2}[K_{0}(F(x))+K_{2}(F(x))]\frac{\partial F(x)}{\partial x}.
\end{equation}
\begin{itemize}
\item For \underline{$\xi=3$}, we have:
\end{itemize}
\begin{equation}
\begin{split}
F^{3}_{d} = - \frac{\partial E^{\xi=3}_{C}}{\partial d} = 
\frac{l^2 \pi^3}{2}\Bigg\{ - \frac{1}{80 d^4} + \frac{9 b 
\zeta(5)}{8 d^5 \pi^4}+ \frac{3}{b \pi^2 d^3}\sum_{n,m=1}^{\infty}\frac{n^2}{m^2} 
K_{2}(2\pi n m b/d) \\
-\frac{3}{2 b d^2 \pi^2}\sum_{n,m=1}^{\infty}\frac{n^2}{m^2} 
\frac{\partial K_{2}(2\pi n m b/d)}{\partial d} \Bigg\} 
\end{split}
\end{equation}
is the force acting on the edges $x=0$ and $x=d$.
\begin{equation}
\begin{split}
F^{3}_{b} = - \frac{\partial E^{\xi=3}_{C}}{\partial b} = 
\frac{l^2 \pi^3}{2}\Bigg\{ - \frac{9 \zeta(5)}{32 d^4 \pi^4} + 
\frac{3}{2 b^2 \pi^2 d^2}\sum_{n,m=1}^{\infty}\frac{n^2}{m^2} K_{2}(2\pi n m b/d) \\
-\frac{3}{2 b d^2 \pi^2}\sum_{n,m=1}^{\infty}\frac{n^2}{m^2} 
\frac{\partial K_{2}(2\pi n m b/d)}{\partial b} \Bigg\} 
\end{split}
\end{equation}
is the force acting on the edges $y=0$ and $y=b$.\\
For this case we have,
\begin{equation}
\frac{\partial K_{2}(F(x))}{\partial x} = -\frac{1}{2}[K_{1}(F(x))+K_{3}(F(x))]\frac{\partial F(x)}{\partial x}.
\end{equation}

\subsubsection*{Neumann}

In this case we must solve \eqref{emret} by imposing the conditions,
\begin{equation}
\begin{cases} 
\frac{\partial\phi}{\partial x}\Big |_{x=0} = \frac{\partial\phi}{\partial x}\Big |_{x=d} = 0\  ,\\
\frac{\partial\phi}{\partial y}\Big |_{y=0} = \frac{\partial\phi}{\partial y}\Big |_{y=b} = 0  \  .
\end{cases}
\end{equation}

Under this condition the corresponding field operator is written by, 
\begin{equation}
\hat{\phi}(x) = \sum_{n,m=0}^{\infty} c_{n,m} \cos\left(\frac{n \pi}{d} x\right) 
\cos\left(\frac{n \pi}{b} y\right)[a_{n,m}e^{-i k_{0}t} + a^{\dagger}_{n,m}e^{i k_{0}t}]\  ,
\end{equation}
where
\begin{equation}
\label{cnm}
c_{n,m} = \begin{cases}
\sqrt{\frac{1}{2 d b k_{0}}}\,\ \mathrm{for} \,\ m=n=0\  , \\ 
\sqrt{\frac{1}{ d b k_{0}}} \,\ \mathrm{for} \,\ m \,\ or \,\ n=0 \\ \  ,
\sqrt{\frac{2}{ d b k_{0}}} \,\ \mathrm{for} \,\ m \,\ \mathrm{and} \,\ n\ge1 \  .
\end{cases}
\end{equation}
Also in this case, we have, $k_{0} = l^{\xi-1}\omega^{\xi}_{n,m}$ and
$\omega_{n,m} = \sqrt{(\frac{n \pi}{d})^2 + (\frac{m \pi}{b})^2}$.

The Hamiltonian operator for this field, is
\begin{equation}
\hat{H} = \frac{b d}{2}\sum_{n,m=0}^{\infty} C^{2}_{n,m} 
 k_{0}^{2}(a^{\dagger}_{n,m}a_{n,m} + 1/2) \  ,
\end{equation}
so that the vacuum energy is given by:
\begin{equation}
\label{ECboxDiric}
E_{0} = \frac{l^{\xi-1}}{4}\left(\frac{\pi}{d}\right)^{\xi}\sum_{n=1}^{\infty} n^{\xi} + 
\frac{l^{\xi-1}}{4}\left(\frac{\pi}{b}\right)^{\xi}\sum_{m=1}^{\infty} m^{\xi} + 
\frac{l^{\xi-1}}{2}\sum_{n,m=1}^{\infty}\omega^{\xi}_{n,m} \   .
\end{equation}

The last term of \eqref{ECboxDiric} is the same vacuum energy for the Dirichlet case 
\eqref{ECDiric}, so  we need only to calculate the first two terms. Using the Abel-Plana 
formula for the sum and results of \cite{Grad}, we get:
\begin{equation}
\label{F1}
\frac{l^{\xi-1}}{4}\left(\frac{\pi}{d}\right)^{\xi}\sum_{n=1}^{\infty} n^{\xi}  = 
\frac{l^{\xi-1}}{4}\left(\frac{\pi}{d}\right)^{\xi}\left\lbrace\int_{0}^{\infty}t^{\xi}\mathrm{d}t - 
2 \sin\left(\xi\frac{\pi}{2}\right)\frac{\Gamma(\xi+1)\zeta(\xi+1)}{(2\pi)^{\xi+1}}\right\rbrace
\end{equation}
and
\begin{equation}
\label{F2}
\frac{l^{\xi-1}}{4}\left(\frac{\pi}{b}\right)^{\xi}\sum_{m=1}^{\infty} m^{\xi}  = 
\frac{l^{\xi-1}}{4}\left(\frac{\pi}{b}\right)^{\xi}\left\lbrace\int_{0}^{\infty}t^{\xi}\mathrm{d}t - 
2 \sin\left(\xi\frac{\pi}{2}\right)\frac{\Gamma(\xi+1)\zeta(\xi+1)}{(2\pi)^{\xi+1}}\right\rbrace \ .
\end{equation}

Then, substituting \eqref{F1}, \eqref{F2} and \eqref{E0RD} into \eqref{ECboxDiric}, we find:
\begin{equation}
\label{E0RN}
\begin{split}
E_{0} = \frac{l^{\xi-1}}{2} & \Bigg\{ \int_{0}^{\infty}\mathrm{d}t\int_{0}^{\infty}\mathrm{d}v 
\left[\left(\frac{v\pi}{d}\right)^2 + \left(\frac{t\pi}{b}\right)^2\right]^{\xi/2} - 
\sin\left(\xi\frac{\pi}{2}\right) \left(\frac{\pi}{b}\right)^\xi \frac{\Gamma(\xi+1)\zeta(\xi+1)}{(2 \pi)^{\xi+1}}  \\
\\
& -  \sin\left(\xi\frac{\pi}{2}\right)\frac{d}{b}\left(\frac{\pi}{b}\right)^\xi 
\frac{\Gamma(\xi+2)\zeta(\xi+2)\mathrm{B}(\frac{1}{2},\frac{\xi}{2}+1)}{(2 \pi d/b)^{\xi+2}} \\ 
 \\
 - 2 & \sin\left(\xi\frac{\pi}{2}\right)\left(\frac{\pi}{d}\right)^\xi 
\frac{\Gamma(\frac{\xi}{2}+1)}{\pi^{\frac{\xi+2}{2}}} \left(\frac{b}{d}\right)^{\frac{1-\xi}{2}}
\sum_{n,m=1}^{\infty} \left(\frac{n}{m}\right)^{\frac{\xi+1}{2}}K_{\frac{\xi+1}{2}}(2 \pi n m b/d)\Bigg\} \ .
\end{split}
\end{equation}

The first integral above refers to the free vacuum energy, as has been said before. 
It is subtracted from the renormalization process. Thus Casimir energy for this case is given by:
\begin{equation}
\begin{split}
E_{C} = -\frac{l^{\xi-1} \pi^\xi }{2} \sin\left(\xi\frac{\pi}{2}\right)\Big\{& 
\frac{\Gamma(\xi+1)\zeta(\xi+1)}{b^\xi (2 \pi)^{\xi+1}} + \frac{b}{d^{\xi+1} 
(2\pi)^{\xi+2}}\Gamma(\xi+2)\zeta(\xi+2)\mathrm{B}\Big(1/2,\frac{\xi}{2} +1\Big) \\ 
&+ 2 \frac{\Gamma(\frac{\xi}{2}+1)}{d^{\frac{\xi+1}{2}} b^{\frac{\xi-1}{2}} 
\pi^{\frac{\xi}{2}+1}}\sum_{n,m=1}^{\infty} \left(\frac{n}{m}\right)^{\frac{\xi+1}{2}}
K_{\frac{\xi+1}{2}}(2 \pi n m b/d)\Big\}   \   .
\end{split}
\end{equation}

Again, for $\xi$ even the Casimir force vanishes. Let us look at the 
two particular cases $\xi=1$ and $\xi=3$:
\begin{itemize}
\item $\xi=1$
\end{itemize}
\begin{equation}
E^{\xi=1}_{C} = -\frac{\pi}{2}\left\lbrace \frac{1}{24 b} + \frac{b}{8 \pi^2 d^2}\zeta(3) + 
\frac{1}{d \pi}\sum_{n,m=1}^{\infty}\frac{n}{m}K_{1}(2\pi n m b/d)\right\rbrace \  .
\end{equation}
\begin{itemize}
\item $\xi=3$
\end{itemize}
\begin{equation}
E^{\xi=3}_{C} =  \frac{l^{2} \pi^{3}}{2}\left\lbrace \frac{1}{240 b^3} + \frac{9 b}{32 \pi^4 d^4}\zeta(5) +
\frac{3 b}{2  d^2 \pi^2}\sum_{n,m=1}^{\infty}\frac{n^2}{m^2}K_{2}(2\pi n m b/d)\right\rbrace \  .
\end{equation}

We can now calculate the Casimir force for these two cases:
\begin{itemize}
\item For \underline{$\xi=1$}:
\end{itemize}
\begin{equation}
\begin{split}
F^{1}_{d} = - \frac{\partial E^{\xi=1}_{C}}{\partial d} = -\frac{\pi}{2}\Bigg\{ \frac{b}{4 \pi^2 d^3}\zeta(3) + 
\frac{1}{\pi d^2}\sum_{n,m=1}^{\infty}\frac{n}{m}K_{1}(2 \pi n m b/d) \\ - \frac{1}{\pi d}
\sum_{n,m=1}^{\infty}\frac{n}{m}\frac{\partial K_{1}(2 \pi n m b/d)}{\partial d}\Bigg\}
\end{split}
\end{equation}
and
\begin{equation}
F^{1}_{b} = - \frac{\partial E^{\xi=1}_{C}}{\partial b} = -\frac{\pi}{2}\left\lbrace \frac{1}{24 b} - 
\frac{\zeta(3)}{8 \pi^2 d^2} - \frac{1}{\pi d}\sum_{n,m=1}^{\infty}\frac{n}{m}
\frac{\partial K_{1}(2 \pi n m b/d)}{\partial b}\right\rbrace,
\end{equation}
where $F^{1}_{d}$ is the force acting on the edges $x=0$, $x=d$ and $F^{1}_{b}$ 
is the force acting on the edges $y=0$, $y=b$. In the above expressions
\begin{equation}
\frac{\partial K_{1}(F(x))}{\partial x} = -\frac{1}{2}[K_{0}(F(x))+
K_{2}(F(x))]\frac{\partial F(x)}{\partial x} \  .
\end{equation}
\begin{itemize}
\item For \underline{$\xi=3$}:
\end{itemize}
\begin{equation}
\begin{split}
F^{3}_{d} = - \frac{\partial E^{\xi=3}_{C}}{\partial d} = 
\frac{l^2 \pi^3}{2}\Bigg\{\frac{9 b \zeta(5)}{8 d^5 \pi^4}+ 
\frac{3b}{ \pi^2 d^3}\sum_{n,m=1}^{\infty}\frac{n^2}{m^2} K_{2}(2\pi n m b/d) \\
-\frac{3b}{2 d^2 \pi^2}\sum_{n,m=1}^{\infty}\frac{n^2}{m^2} 
\frac{\partial K_{2}(2\pi n m b/d)}{\partial d} \Bigg\} 
\end{split}
\end{equation}
is the force acting on the edges $x=0$, $x=d$, and 
\begin{equation}
\begin{split}
F^{3}_{b} = - \frac{\partial E^{\xi=3}_{C}}{\partial b} = \frac{l^2 \pi^3}{2}\Bigg\{\frac{1}{80 b^4}  - 
\frac{9 \zeta(5)}{32 d^4 \pi^4} - \frac{3}{2 \pi^2 d^2}\sum_{n,m=1}^{\infty}\frac{n^2}{m^2} K_{2}(2\pi n m b/d) \\
-\frac{3 b}{2  d^2 \pi^2}\sum_{n,m=1}^{\infty}\frac{n^2}{m^2} \frac{\partial K_{2}(2\pi n m b/d)}{\partial b} \Bigg\}
\end{split}
\end{equation}
is the force acting on the edges $y=0$, $y=b$. Also we have
\begin{equation}
\frac{\partial K_{2}(F(x))}{\partial x} = -
\frac{1}{2}[K_{1}(F(x))+K_{3}(F(x))]\frac{\partial F(x)}{\partial x}.
\end{equation}

\subsubsection*{Mixed Condition}

We impose mixed boundary conditions to solve \eqref{emret}:
\begin{equation}
\begin{split}
i) \begin{cases} \,\ \phi(x=0) = \frac{\partial\phi}{\partial x} |_{x=d}=0 \ , \\  
\phi(y=0) = \frac{\partial\phi}{\partial y} |_{y=b}=0 \end{cases} \ . \\ 
ii) \begin{cases} \,\ \frac{\partial\phi}{\partial x} |_{x=0}= \phi(x=d) = 0 \ , \\   
\frac{\partial\phi}{\partial y} |_{y=0}= \phi(y=b) =0 \end{cases}\ .
\end{split}
\end{equation}

The field operators consistent with these boundary conditions are given by,
\begin{equation}
\hat{\phi}_{i}(x) = \sum_{n,m=o}^{\infty} \sqrt{\frac{2}{d b k_{0}}} 
\mathrm{sin}\left[(n+1/2)\frac{\pi}{d} x\right] \mathrm{\sin}
\left[(m+1/2)\frac{\pi}{b} y\right][a_{n,m}e^{-i k_{0}t} + a^{\dagger}_{n,m}e^{i k_{0}t}]
\end{equation}
and
\begin{equation}
\hat{\phi}_{ii}(x) = \sum_{n,m=o}^{\infty} \sqrt{\frac{2}{d b k_{0}}} 
\cos\left[(n+1/2)\frac{\pi}{d} x\right] \cos\left[(m+1/2)\frac{\pi}{b} y\right]
[a_{n,m}e^{-i k_{0}t} + a^{\dagger}_{n,m}e^{i k_{0}t}] \  ,
\end{equation}
where $\omega_{n,m} = \sqrt{\left[(n+1/2)\frac{\pi}{d}\right]^2 + 
\left[(m+1/2)\frac{\pi}{b}\right]^2}$.

The Hamiltonian operators obtained for both fields are the same:
\begin{equation}
\hat{H} = \frac{l^{\xi-1}}{2}\sum_{n,m=0}^{\infty}\omega^{\xi}_{n,m}[2a^{\dagger}_{n,m}a_{n,m} + 1].
\end{equation}

So the vacuum energy is given by:
\begin{equation}
E_{0} = \frac{l^{\xi-1}}{2}\pi^{\xi}\sum_{n,m=0}^{\infty} 
\left[\left(\frac{n+1/2}{d}\right)^2 + \left(\frac{m+1/2}{b}\right)^2\right]^{\xi/2}.
\end{equation}

Using the Abel-Plana formula and analyzing the integration interval in the second term of the
sum, as it has been done several times during this work, we get:
\begin{equation}
\begin{split}
\label{ECBoxMix}
E_{0} = \frac{l^{\xi-1}}{2}\pi^{\xi}\sum_{n=0}^{\infty} 
\Bigg\{ \underbrace{\int_{0}^{\infty}\mathrm{d}t\left[\left(\frac{t}{b}\right)^2 + 
\left(\frac{n+1/2}{d}\right)^2\right]^{\xi/2}}_{I} +\\ \underbrace{2 \sin\left(\xi\frac{\pi}{2}\right)
 \int_{(n+1/2)b/d}^{\infty}\mathrm{d}t\dfrac{[\left(\frac{t}{b}\right)^2 - 
\left(\frac{n+1/2}{d}\right)^2]^{\xi/2}}{e^{2 \pi t}+1}}_{II}\Bigg\}\   .
\end{split}
\end{equation}

Using the Abel-Plana formula to perform the sum of $n$ for term $I$ of \eqref{ECBoxMix}, we have:
\begin{equation}
I = \int_{0}^{\infty}\mathrm{d}t\int_{0}^{\infty}\mathrm{d}v\left[\left(\frac{t}{b}\right)^2 + 
\left(\frac{v}{d}\right)^2\right]^{\xi/2} + 2 \sin\left(\xi\frac{\pi}{2}\right) 
\int_{0}^{\infty}\mathrm{d}t\int_{td/b}^{\infty}\mathrm{d}t\dfrac{[\left(\frac{v}{d}\right)^2 - 
\left(\frac{t}{b}\right)^2]^{\xi/2}}{e^{2 \pi t}+1} \   .
\end{equation}

The integral of the second term of $I$ is obtained from \cite{Grad}, resulting in:
\begin{equation}
\begin{split}
\label{F1Mix}
I = &\int_{0}^{\infty}\mathrm{d}t\int_{0}^{\infty}\mathrm{d}v\left[\left(\frac{t}{b}\right)^2 + 
\left(\frac{v}{d}\right)^2\right]^{\xi/2} + \\
&(1-2^{-(\xi-1)})\frac{b}{d^{\xi+1}} \sin\left(\xi\frac{\pi}{2}\right) 
\frac{B \Big(1/2, \frac{\xi}{2}+1\Big)\Gamma(\xi+2)\zeta(\xi+2)}{(2\pi)^{\xi+2}} \  .
\end{split}
\end{equation}
 
The term $II$ of \eqref{ECBoxMix} is given by:
\begin{equation}
II = 2 \sin\left(\xi\frac{\pi}{2}\right) \sum_{n=0}^{\infty}\int_{(n+1/2)b/d}^{\infty}
\mathrm{d}t\dfrac{[\left(\frac{t}{b}\right)^2 - \left(\frac{n+1/2}{d}\right)^2]^{\xi/2}}{e^{2 \pi t}+1}.
\end{equation}
Using:
\[\frac{1}{e^{m}+1} = -\sum_{l=1}^{\infty}(-1)^le^{-nl},\]
we can rewrite $II$ as:
\begin{equation}
II = - 2 \sin\left(\xi\frac{\pi}{2}\right) \sum_{n=0}^{\infty}\sum_{m=1}^{\infty}(-1)^m
\int_{(n+1/2)b/d}^{\infty}\mathrm{d}t\left[\left(\frac{t}{b}\right)^2 - 
\left(\frac{n+1/2}{d}\right)^2\right]^{\xi/2}e^{-2 \pi t m }\  .
\end{equation}

Finally by using again \cite{Grad}, we find:
\begin{equation}
\begin{split}
\label{F2Mix}
II = - 2 \sin\left(\xi\frac{\pi}{2}\right)& \frac{b^{\frac{1-\xi}{2}}}
{d^{\frac{\xi+1}{2}}}\frac{\Gamma(\frac{\xi}{2}+1)}{\pi^{\frac{\xi}{2}+1}} 
\sum_{n=0}^{\infty}\sum_{m=1}^{\infty}(-1)^m \left(\frac{n+1/2}{m}\right)^{\frac{\xi+1}{2}}
\\ & \times K_{\frac{\xi+1}{2}}(2 \pi(n+1/2)m b/d)\  .
\end{split}
\end{equation}

Substituting the results \eqref{F1Mix} and \eqref{F2Mix} into \eqref{ECBoxMix}, we have:
\begin{equation}
\begin{split}
E_{0} = \frac{l^{\xi-1}\pi^\xi}{2} \Bigg\{ &\int_{0}^{\infty}\mathrm{d}t
\int_{0}^{\infty}\mathrm{d}v\left[\left(\frac{t}{b}\right)^2 + 
\left(\frac{v}{d}\right)^2\right]^{\xi/2} + \\
&(1-2^{-(\xi-1)})\frac{b}{d^{\xi+1}} \sin\left(\xi\frac{\pi}{2}\right) 
\frac{B \Big(1/2, \frac{\xi}{2}+1\Big)\Gamma(\xi+2)\zeta(\xi+2)}{(2\pi)^{\xi+2}}   \\
 -&2 \sin\left(\xi\frac{\pi}{2}\right)  \frac{b^{\frac{1-\xi}{2}}}{d^{\frac{\xi+1}{2}}}
\frac{\Gamma(\frac{\xi}{2}+1)}{\pi^{\frac{\xi}{2}+1}} \sum_{n=0}^{\infty}
\sum_{m=1}^{\infty}(-1)^m \left(\frac{n+1/2}{m}\right)^{\frac{\xi+1}{2}}
\\ &\times K_{\frac{\xi+1}{2}}(2 \pi(n+1/2)m b/d)\Bigg\} \ .
\end{split}
\end{equation}

Again, the first integral refers to the free vacuum energy in an area bounded 
by the "rectangle" $ab$, however without the contours, so this term is subtracted 
from the renormalization process. Consequently, the Casimir energy for this boundary condition is given by:
\begin{equation}
\begin{split}
\label{ECMix}
E_{C} = \frac{l^{\xi-1}\pi^\xi}{2}\sin\left(\xi\frac{\pi}{2}\right) \Bigg\{&(1-2^{-(\xi-1)})
\frac{b}{d^{\xi+1}} \frac{B \Big(1/2, \frac{\xi}{2}+1\Big)\Gamma(\xi+2)\zeta(\xi+2)}{(2\pi)^{\xi+2}} \\
- \,\ 2 \frac{b^{\frac{1-\xi}{2}}}{d^{\frac{\xi+1}{2}}}\frac{\Gamma(\frac{\xi}{2}+1)}
{\pi^{\frac{\xi}{2}+1}}& \sum_{n=0}^{\infty}\sum_{m=1}^{\infty}(-1)^m 
\Big(\frac{n+1/2}{m}\Big)^{\frac{\xi+1}{2}}K_{\frac{\xi+1}{2}}(2 \pi(n+1/2)m b/d)\Bigg\}. 
\end{split}
\end{equation}

As we see again, for $\xi$ even the Casimir energy is zero. So, let us present two particular cases, for
 $\xi=1$ and $\xi=3$:
\begin{itemize}
\item $\xi=1$
\end{itemize}
\begin{equation}
E^{\xi=1}_{C} = \frac{\pi}{2}\left\lbrace \frac{3}{32}\frac{ b \zeta(3)}{d^2 \pi^2} - 
\frac{1}{d\pi}\sum_{n=0}^{\infty}\sum_{m=1}^{\infty}(-1)^m \Big(\frac{n+1/2}{m}\Big)
K_{1}(2 \pi(n+1/2)m b/d)\right\rbrace \  .
\end{equation}
\begin{itemize}
\item $\xi=3$
\end{itemize}
\begin{equation}
E^{\xi=3}_{C} = - \frac{l^{2}\pi^{3}}{2}\left\lbrace \frac{135}{512}
\frac{ b \zeta(5)}{d^4 \pi^4} - \frac{3}{2 b d^2\pi^2}\sum_{n=0}^{\infty}
\sum_{m=1}^{\infty}(-1)^m \Big(\frac{n+1/2}{m}\Big)^2K_{2}(2 \pi(n+1/2)m b/d)\right\rbrace \  .
\end{equation}

We can now calculate the Casimir forces:
\begin{itemize}
\item For $\xi=1$: 
\end{itemize}
\begin{equation}
\begin{split}
F^{1}_{d} = -\frac{\pi}{2}\Big\{ - \frac{3}{16}\frac{ b \zeta(3)}{d^3 \pi^2} + 
\frac{1}{d^2\pi}\sum_{n=0}^{\infty}\sum_{m=1}^{\infty}(-1)^m \Big(\frac{n+1/2}{m}\Big)
K_{1}(2 \pi(n+1/2)m b/d)  \\ - \frac{1}{d\pi}\sum_{n=0}^{\infty}
\sum_{m=1}^{\infty}(-1)^m \Big(\frac{n+1/2}{m}\Big)\frac{\partial K_{1}(2 \pi(n+1/2)m b/d)}{\partial d} \Big\} \  ,
\end{split}
\end{equation}
that is the force acting on the edges $x=0$, $x=d$, and
\begin{equation}
F^{1}_{b} = -\frac{\pi}{2}\left\lbrace \frac{3}{32}\frac{ \zeta(3)}{d^2 \pi^2} - 
\frac{1}{d\pi}\sum_{n=0}^{\infty}\sum_{m=1}^{\infty}(-1)^m \Big(\frac{n+1/2}{m}\Big)
\frac{\partial K_{1}(2 \pi(n+1/2)m b/d)}{\partial b}\right\rbrace \ ,
\end{equation}
that is the force acting on the edges $y=0$, $y=b$.

In the above expressions, we can write,
\begin{eqnarray}
\frac{\partial K_{1}(F(x))}{\partial x} = -\frac{1}{2}[K_{0}(F(x))+K_{2}(F(x))]\frac{\partial F(x)}{\partial x} \  .
\end{eqnarray}
\begin{itemize}
\item For $\xi=3$: 
\end{itemize}
\begin{equation}
\begin{split}
F^{3}_{d} = \frac{l^2\pi^3}{2}\Big\{ - \frac{135}{128}\frac{ b \zeta(5)}{d^5 \pi^4} + 
\frac{3}{b d^3\pi^2}\sum_{n=0}^{\infty}\sum_{m=1}^{\infty}(-1)^m \Big(\frac{n+1/2}{m}\Big)^2 
K_{2}(2 \pi(n+1/2)m b/d)  \\ - \frac{3}{2 b d^2\pi2}\sum_{n=0}^{\infty}
\sum_{m=1}^{\infty}(-1)^m \Big(\frac{n+1/2}{m}\Big)^2\frac{\partial K_{2}(2 \pi(n+1/2)m b/d)}{\partial d} \Big\} \  .
\end{split}
\end{equation}
and
\begin{equation}
\begin{split}
F^{3}_{b} = \frac{l^2\pi^3}{2}\Big\{ \frac{135}{512}\frac{ \zeta(5)}{d^4 \pi^4} + \frac{3}{2 b^2 d^2\pi^2}
\sum_{n=0}^{\infty}\sum_{m=1}^{\infty}(-1)^m \Big(\frac{n+1/2}{m}\Big)^2\ K_{2}(2 \pi(n+1/2)m b/d) \\ - 
\frac{3}{2 b d^2\pi2}\sum_{n=0}^{\infty}\sum_{m=1}^{\infty}(-1)^m \Big(\frac{n+1/2}{m}\Big)^2\frac{\partial 
K_{2}(2 \pi(n+1/2)m b/d)}{\partial b}\Big\} \  .
\end{split}
\end{equation}
These forces act on the edges, $x=0$, $x=d$, and $y=0$, $y=b$, respectively. Moreover, we can write, 
\begin{equation}
\frac{\partial K_{2}(F(x))}{\partial x} = -\frac{1}{2}[K_{1}(F(x))+K_{3}(F(x))]\frac{\partial F(x)}{\partial x} \ .
\end{equation}

We have seen then that the Lorentz symmetry breaking, employed by HL theory, modifies the
vacuum structure, resulting in a change in the Casimir effects results. The HL theory parameter $\xi$
is the one that dictates what the outcome of the effect. We have seen that for $\xi$ even, the Casimir force is always
zero, while for $\xi$ odd the Casimir force switches between an attractive and a repulsive ones.

\section{Concluding Remarks}
\label{Concl}

In this paper we have investigated the Casimir effects associated to a massless scalar 
real quantum field in the theory with a space-time asymmetry. To do it, we used the 
modified Klein-Gordon equation to study the Casimir effect in the HL-like theory.
Two different situations are considered: The field confined between two 
parallel plates, and the field confined in a two-dimensional rectangular boxes.
In this case we have admitted that the field obeys specific boundary condition.

First we study the Casimir effect for two parallel plates of area $L^2$ separated by a distance $d$. 
For this case, we imposed three different types of boundary conditions on field $\phi(x)$ at 
the boundary: Dirichlet, Neumann and mixed conditions. In each case we obtained the Casimir energy, 
and found that in all cases, it is given by a divergent sum. We evaluated this sum by
using the Abel-Plana summation formulas. Considering the Dirichlet and Neumann conditions, 
we observed that this sum has a three-term contribution: the free vacuum energy contribution
(borderless), vacuum energy contribution in the presence of only one plate, and finally the 
vacuum energy contribution in the presence of two plates. At the same time, for the mixed conditions, 
we saw that the sum is contributed by only two terms: vacuum energy contribution to the presence of 
only one plate and the vacuum energy contribution in the presence of two plates. 
Both contributions, free vacuum energy and the vacuum energy in the presence of only one plate 
are infinite terms that are subtracted by the renormalization process, thus we obtained the 
finite Casimir energy. After that we calculate the Casimir pressures. As in the usual 
cases where the Lorentz symmetry is preserved, the Casimir pressure to the Dirichlet and Neumann 
conditions of are equal and differ from the one to the mixed condition. In all three 
cases we have seen that the Casimir pressure depends on the HL theory parameter $\xi$. 
For $\xi=1$ we recover the usual results where the Lorentz symmetry is preserved, 
for $\xi$ even we saw that the Casimir pressure is always zero, while for 
$\xi$ odd the Casimir pressure switches corresponding to an attractive or repulsive forces. 

We also consider the Casimir effect in a two-dimensional rectangular box with edges 
“$b$” and “$d$”. Again we impose three different types of boundary conditions to the 
field $\phi(x)$: Dirichlet, Neumann and mixed one. We found that the Casimir energy 
for the three cases is given in terms of infinite sums. Using the Abel-Plana formula 
to develop these sums, we saw that the Casimir energy is again given by three 
contributions: the free vacuum energy, the vacuum energy at the presence only two edges 
of the rectangle and the vacuum energy in the presence of the rectangle. We have 
seen that the free vacuum energy and the vacuum energy at presence of only two edges 
of the rectangle are terms infinite, and these are subtracted in the renormalization 
process, so that we obtained the finite Casimir energy. Thus we calculate the Casimir forces and 
see that they depend on the HL theory parameter $\xi$, where for $\xi$ even the 
forces vanishes, while for $\xi$ odd, we have the forces that switches the signal,
which can be an attractive or repulsive force depending on the value of $\xi$.

In general, we can affirm that the HL-like modification change 
the equation of motion that a field must satisfy in a theory, and this change has a crucial 
implication on a very well known effect in the literature, the Casimir one.
Moreover, the Casimir force also depends also on the types of conditions
and also depend on HL theory parameter $\xi$.
 
Where once the Casimir force depended only on the type of condition imposed on the 
field, we have that in this context the Casimir force also depends on a theory parameter of in question.

The natural application of the results we obtained can be the following one. Since the Casimir 
effect now can be measured in a very precise manner, our corrections to the Casimir effect arising 
from the Lorentz-breaking modifications of the field theory action, calculated theoretically 
and then compared with the experiment, can serve for estimating the values of the Lorentz-breaking 
parameters in the corresponding theory.

A natural continuation of this study could consist, first, in consideration of Casimir effect for other
fields; second, in studying of the Casimir effect for small Lorentz-breaking additive
corrections whose actions are discussed in \cite{Carroll}, \cite{Kost} and \cite{Alfaro}.

\section*{Acknowledgment} 
IJMU thanks the Coordenação de Aperfeiçoamento de Pessoal de Nível Superior (CAPES). ERBM and
AYP  thank Conselho Nacional de Desenvolvimento Cient\'\i fico e Tecnol\'ogico (CNPq).

\end{document}